# GUIDELINES FOR A DYNAMIC ONTOLOGY

*Integrating Tools of Evolution and Versionning in Ontology*

Perrine Pittet, Christophe Cruz, Christophe Nicolle
*LE2I, UMR CNRS 5158*
*University of Bourgogne - Dijon, France*
*{perrine.pittet, christophe.cruz, christophe.nicolle}@u-bourgogne.fr*

Abstract: Ontologies are built on systems that conceptually evolve over time. In addition, techniques and languages for building ontologies evolve too. This has led to numerous studies in the field of ontology versioning and ontology evolution. This paper presents a new way to manage the lifecycle of an ontology incorporating both versioning tools and evolution process. This solution, called VersionGraph, is integrated in the source ontology since its creation in order to make it possible to evolve and to be versioned. Change management is strongly related to the model in which the ontology is represented. Therefore, we focus on the OWL language in order to take into account the impact of the changes on the logical consistency of the ontology like specified in OWL DL.

## 1 INTRODUCTION

According to (Hodgson, 2003), ontology lifecycle is divided in seven steps: needs detection, conception, management and planning, evolution, diffusion, use, and evaluation. The needs detection phase starts with a detailed inventory of the domain and the various purposes. Like evolution phase, conception phase needs: knowledge acquisition, shared conceptualization building, formalization (Semantic Web [1] formalisms…) and integration of the existing resources (another ontology, applications…).The The phase of management and planning underlines the importance of having a constant monitoring and a global policy to detect or initiate, prepare or evaluate the lifecycle iterations. This work intends to guarantee that an iteration of the lifecycle is activated when an evolution is ready to be completed. The management step requires tools not only to prepare the ontology to adapt the domain changes but also to keep tracing of the previous versions of the ontology. These goals can be reached with a versioning system (Flouris and al, 2007). Diffusion phase deals with the deployment of the ontology. The use phase encloses all the activities related to the access of the ontology. Finally, the evaluation phase aims at evaluating the ontology state. Moreover, like the needs detection phase, it collects beforehand the knowledge of the domain and can also rely on previous studies or feedbacks. Except for the evolution and management phases, all the steps described can be considered as mature domains. Furthermore, this description of the lifecycle shows that evolution, and management remains the most complex phases. Evolution is the backbone of the lifecycle iterations. Therefore, the change management process is totally based on it. Our state of art is articulated in three parts. According to the literature, we will first define the evolution role, operations and process. Then we'll have a look at the existing solutions for change representation and ontology versioning. We will see how to link the evolution process and a versioning system in order to integrate both in existing ontologies.

## 2 ONTOLOGY EVOLUTION

As stated by (Flouris and al, 2007), ontology evolution aims at responding to one or several changes in the domain or the conceptualization by applying them on the source ontology. This brief definition looks abstract and leads us to ask: what kind of changes does the evolution apply? How evolution applies them? What are the criteria to respect? How can we manage a good evolution? Evolution changes are defined in the literature and especially in (Noy and Klein, 2004) as a succession of simple or complex operations the user wants to apply on the intension (schema) or the extension (data) of the ontology. This evolution aims at adapting the ontology to the changed domain. Applying and propagating the change are often manual tasks but can be done automatically by synchronization with the domain. According to (Tovar, and Vidal, 2008) these tasks usually occur during the use phase of the ontology. Ontology Dynamics clearly define the evolution criteria. (Atle and Sugumaran, 2008) and (Dividino and Sonntag, 2008) qualify the maintenance of the ontology as the most important criterion. Evolution has to maintain whatever relies on the ontology. Maintaining the ontology consistent and pertinent, in a consensus is an inescapable issue of evolution (Zablith and al, 2008). Applying changes on ontology can turn the conceptualization inconsistent and irrelevant. That's why an evolution should never be validated before the user has a preview of the impact of the changes on the ontology. This impact can only be estimated if the evolution operations are semantically clearly defined. In order to assure that this process is fully respected, some works propose an approach in six phases.

---

[1] Semantic Web: http://semanticweb.org/wiki/Main_Page

1. The **change detection** phase consists in detecting what changes occurred in the domain or in the point of view must be propagated to the conceptualization. Lots of papers in the Ontology Dynamics deal with this phase and propose methods and tools like integrated event handlers (Tovar and Vidal, 2008), ontology learning (Novacek and al) etc.

2. The **representation phase** aims at representing the selected changes with ontological operations. (Noy and Klein, 2004) classifies the evolution operations in two types: elementary (atomic) operations and composed (complex) operations. According to (Noy and Klein, 2004), elementary operations are simple operations that modify only one entity like addition/suppression of classes/relations, of hierarchy, domain, range links, of class/relation properties like disjoint, transitivity, etc…whereas composed operations are a composition of several elementary operations. The choice of composed operations depends on the granularity of the evolution needs. Usual operations correspond to operations the ontology that developers are the most expected to use when creating and evolving an ontology. In addition to elementary operations, the literature gives some lists of usual operations (Stojanovic and al, 2002,Stickenschmidt and Klein, 2003). A distinction can be done between operations on the intension and operations on the extension. The cited works on change operations do not specify specific operations for the instances because they argue that an instance can become a class (Noy and Klein, 2004). However, we maintain that schema operations can't be confounded with instance operations. Actually, it is impossible to create an instance (instance operation) related to a class if this class is not created. Inversely a class can be created (schema operation) without instances.

3. The **semantic phase** prevents the user from inconsistency risks by determining the sense of the represented changes. For example, if composed operations have been selected, this phase will allow seeing their decomposition in elementary operations.

4. The **implementation** of the changes alerts the user of the impact on data in terms of data gain or loss. (Noy and Klein, 2004) gives these impacts from a list of 22 usual operations (the elementary ones and some composed).

5. The **propagation phase** aims at informing all the dependent parts of the ontology (other ontologies, application) of these changes.

6. Finally, in sixth step comes the **validation** of the changes.

## 3 ONTOLOGY VERSIONING

This part defines the role of versioning, bringing our new vision on this definition. First, (Flouris and al, 2007) gives in 2007 a very strict definition of the role of versioning: give a transparent access to different existing versions of an ontology by creating a versioning system. This system identifies the versions by their "Id" and delimits their mutual compatibility. In the past three years, Ontology Dynamics proposals extend its role: manage several chronological and multitemporal versions (Grandi, 2008), at a local or web level (Allocca and al), when collected, distributed, accessed by search engines. All these definitions correspond to a retroactive versioning because versions of the ontology have to preexist. However, in our objective, we want to integrate a versioning system since the creation of the first version of the ontology, and we want it to be reactive when a change occurs. Therefore, we need, as the ontology development, a dynamic and incremental process, which could take into account a new version at each evolution phase. That is why we propose to merge the evolution process (following the six phases) with the versioning one. (Sassi and al, 2010) and (Djedidi and Aufaure, 2008) agree with this proposition by giving the ontology versioning the ability of following the evolution process. In and (Djedidi and Aufaure, 2008), the methodology goal is to guide and validate the application of the changes in a systematic and optimized way, maintaining the coherence and evaluating the impact of the change on the ontology quality by the mean of design patterns. In (Sassi and al, 2010), the goal is to assist the users during the evolution process to observe the consequences of the change applications on the several versions by allowing them to compare them. The two methodologies are step by step approaches integrating the versioning process directly into the evolution one. Both propositions quite follow the evolution phases cited before] but do not explicitly show them.

## 4 VERSIONGRAPH APPROACH

This section presents the versioning approach of our versioning system based on the six phases of the evolution process.

### 4.1 From Evolution Phases to Versioning

To make sure the evolution phases are fully respected we chose to match each of them with a versioning step. First, the user chooses the list of operations to apply: (cf. change detection phase). The versioning system formalizes them (cf. representation phase), turn them semantically understandable (cf. semantic phase), records and implements them (cf. implementation phase). Then after the propagation of the changes, (cf. propagation phase), the user validates them (cf. validation phase) and the versioning system applies them and generates the new version of the ontology corresponding to an evolution iteration. Finally, the versioning system can give a transparent access to both versions with criteria defined by the user (Stuckenschmidt and Klein, 2003). It can delimit compatibility by retracing evolution operations (Stojanovic and al, 2002, Stuckenschmidt and Klein, 2003).

### 4.2 Versioning Steps Tools

To follow this process, we need to specify the tools displayed by our versioning system. According to (Klein and Fensel, 2001), a change specification should enclose an operational change specification (our list of operations), next the conceptual relationship between the first version and the new one (the selected operations on the selected entities). The first phase of the evolution process is then completed. The next step is to represent these changes. Several approaches are proposed in the literature to represent changes. Major part of them uses logs. Versioning logs (Yildiz, 2005) record the different versions of an ontology by representing each entity at a given time. For each class, relation and instance, a new instance of "EvolutionConcept" class is created. (Klein and Fensel, 2001) argues that metadata should be added to identify this change. In versioning logs, each instance is annotated with metadata (Id, cause, transaction time, state validated or not, etc.). This solution is interesting if the versioning log can be integrated in the ontology. However, for our purposes, there is no need to represent each entity if it is not modified by the evolution. Evolution logs (Liang, 2005) do not save the versions but act like a change history. Not each entity but each substitution in the ontology is recorded in order to be reused when the user wants to access a version. Tracing the substitution rather corresponds to our objectives as a substitution contains the selected operations and the entities affected. In order to cope with our evolution process, we propose to create a Version concept like in the versioning logs integrated in the ontology that will be created at each evolution iteration. This Version concept encloses: 1/the substitutions operated in the intension or 2/ those operated on the extension and 3/ the metadata. For the semantic phase, we chose to use ontology design patterns (ODP) (Gangemi, 2005)) as (Djedidi and Aufaure, 2008) proposes in addition to an evolution log, in order to guarantee the consistence of the ontology when applying the change. Then, the implementation phase can be helped by introducing event detectors on data. In the Jena application supporting the ontology, the idea is to insert methods using "ActionListener" objects. The propagation phase can be performed by generating events activating the "ActionListener" objects. Finally, the validation is similar to the "Commit" operator of a DBMS, can be done by a simple click by the user. Our incremental versioning process following the six evolution phases constitutes the first part of our versioning system.

### 4.3 Version Retrieval

Concerning the transparent access definition, the first issue is the identification of the versions. Most of the versioning systems use "Id" of the ontologies to identify them (Allocca and al, 2008). Though, it is not enough to identify in which version a change on a certain entity occurred. As we have introduced the metadata and the list of substitutions occurred when a Version is created, those data can serve as search criteria to identify and retrieve the right version. We have chosen to extend Jena's operators (access on ontology, etc.) in order to take into account the search criteria. This extension can be performed by an override of the access methods, for example, by adding metadata and operation attributes. This state of art permitted us to build the evolution and versioning process of our proposition. We also managed to design the versioning tools in order to represent changes and access the ontology.

## 5 VERSIONGRAPH ARCHITECTURE

In this section, we present the VersionGraph architecture which implements the choices of our state of art.

### 5.1 Evolution Operations

Contrarily, to the (Sassi and al, 2010) proposition, the schema and instance operations are differentiated respectively by `SchemaOperation` and `InstanceOperation`. `SchemaOperation` type operations correspond to the creation and deletion of classes (`AddClass`) and properties (`AddProperty`) but also to additions and deletions of restrictions on them. We distinguish restrictions on the classes and properties or properties of the data link hierarchy (`HierarchyLink`) such as class / subclass, property / sub-property. Furthermore, in the class restrictions, limitations like classes / properties such as the relationship between properties and classes (`ClassPropertyLink`, `ClassDataPropertyLink`), car-dinality (`ClassPropertyCardinality`) are classified. In addition, in the restrictions we find domain and range restrictions of attributes (`PropertyAttributeLink`). Finally, `TypeProperty` operations are used to define a specific constraint of a property (transitive, symmetric, etc.).

`InstanceOperation` type operations correspond to operations of addition and deletion of individuals and statements about these individuals. We distinguish between the assertions relying individuals to the values (`DataPropertyAssertion`) and those specifying the types for these individuals (`ObjectPropertyAssertion`).

## 5.2 Versioning Process

From these evolution operations and the study of the different versioning solutions of our state of art, we derived a versioning system. At each evolution of the ontology, the system stores in the ontology, the changes impacted by the operations used and the context. This versioning system is an independent ontology which intends to be integrated into the existing ontology by a simple addition operation. Then, the user can start a first evolution of ontology in choosing whether to change the schema (intension) or data (extension) using the above operations. Each list of changes chosen by the user during the evolution is kept using a concept `SchemaVersionGraph` for `SchemaOperation` operations and `InstanceVersionGraph` for `Instance-Operation` operations on instances by specifying which elements of the ontology are concerned (concepts, relationships, etc.). Contextual information can be added (as version, date, author, description, etc.). These data are traced during the evolution using a concept of context `VersionContext`. The set containing `SchemaVersion-Graph` or `InstanceversionGraph` and `Version-Context` is called `VersionGraph`. Figure 1 depicts an overview of the ontology schema. For more clarity, it only shows concepts and their relationships under $6^{th}$ hierarchical degrees.

In a transparent way, each application of changes made by the user generates a new `VersionGraph`. A `VersionGraph` contains a link with the previous version of the ontology (`hasPrevious-VersionGraph`). It's actually a link to the core ontology (for the first `VersionGraph`) or to the previous `VersionGraph`. Because of its nature, our system of evolution and versioning can be integrated into applications using ontologies Jena. The access operations of the library Jena can be overridden by the criteria of change and context. Until now, proposals for versioning are often accompanied by a specific application that the user must install to access the version it wants if the use of URI is not enough (Evolva). However, many ontologies are accessed using a Java API Jena. Indeed, this library supports ontology-based formalisms like RDF, RDFS, OWL and the various DAML + OIL. Jena contains all the methods to access and edit ontologies. In addition, it also implements all the basic operations of evolution and the commonly used composed ones. Overridden access methods are able to take into account the criteria of versions thanks to new attributes. These criteria are integrated into the ontology itself as we saw in the previous paragraph.

## 4.3 The Wine Ontology Versionning

International wines are described at `<http://www.w3.org/TR/owl-guide/wine.rdf>`;
Afterwards, we want to add the "StrawWine" wine which does not exist in the Wine ontology. Straw Wine's fruit is selected then dried in the sun so that the juice is very concentrated in flavor and sugar. Consequently, it is a dessert style wine sometimes heavy or balanced or straw gold color. It can be made from red grapes Cabernet Franc and Cabernet Sauvignon or Chardonnay white grapes and Sauvignon Blanc. To add this new concept and describe it, the system creates another `VersionGraph`. This new one is linked with the previous one. The system specifies a SchemaVersionGraph which contains the operations needed to describe and add the concept in the ontology.

The Wine ontology is an ontology example in which international wines are described. For the first step, the VersionGraph ontology is imported into the Wine ontology by an addition operation (Script 1). Then the system creates the first version of the wine ontology with a primary instance of `VersionGraph`. This Versiongraph only has a link with the source ontology. Next, we want to add the "StrawWine" wine which doesn't exist in the Wine ontology.

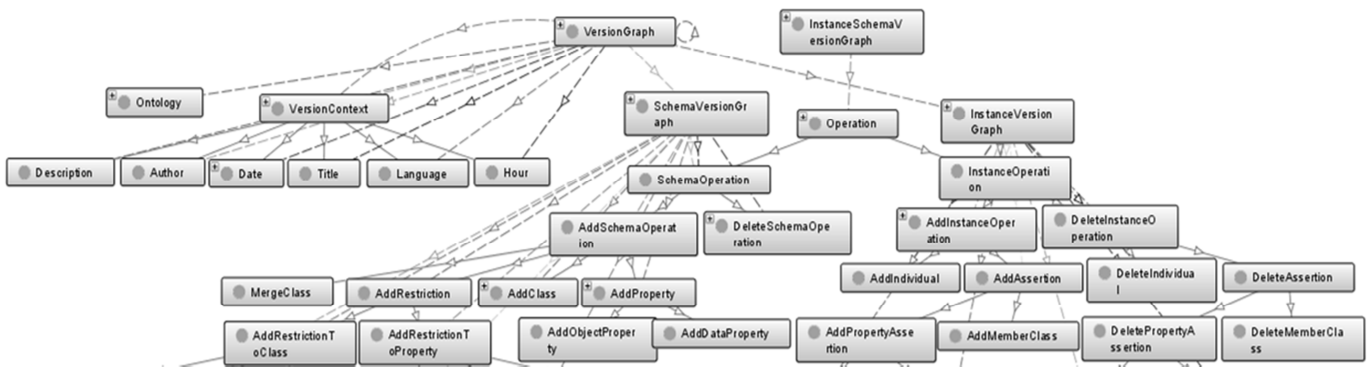

Figure 1. VersionGraph definition in Protege.

Straw Wine's fruit is selected then dried in the sun so that the juice is very concentrated in flavor and sugar. So it is a dessert style wine sometimes heavy or balanced or straw gold color. It can be made from red grapes Cabernet Franc and Cabernet Sauvignon or Chardonnay white grapes and Sauvignon Blanc. To add this new concept and describe it, the system creates another VersionGraph. This new one is linked with the previous one. The system specifies a SchemaVersionGraph which contains the operations needed to describe and add the concept in the ontology (Script 2).

Script 1. Version graph for the Wine ontology

```
<vg :VersionGraph#VersionGraph0>
    p:hasPreviousVersionGraph    <http://www.w3.org/TR/owl-guide/wine.rdf>;
```

Script 2. Version graph extended with new instances.

```
# VersionGraph1 description
<vg:VersionGraph#VersionGraph1>
    p:hasPreviousVersionGraph    <vg:VersionGraph#VersionGraph0>;
    p:hasDate                    "11/05/2010";
    p:hasAuthor                  "Perrine PITTET";
    p:hasSchemaVersionGraph      <vg:SchemaVersionGraph#SchemaVersionGraph1>;

# AssociatedSchemaVersionGraph1 description
<vg:SchemaVersionGraph#SchemaVersionGraph1>
    p:hasAddClass                <rdfs:class#StrawWine>;
    p:hasAddClassHierarchyLink   <vg:ClassHierarchyLink#ClassHierarchyLink1>;
    p:hasAddClassDataPropertyLink <vg:ClassDataPropertyLink#ClassDataPropertyLink1>;
    p:hasAddClassDataPropertyCardinality
                <vg:ClassDataPropertyCardinality#ClassDataPropertyCardinality1>;
    p:hasAddClassDataPropertyCardinality
                <vg:ClassDataPropertyCardinality#ClassDataPropertyCardinality2>;

# Description of SchemaOperation used
<vg:ClassHierarchyLink#ClassHierarchyLink1>
    p:class                      <rdfs:class#StrawWine>;
    p:subClass                   <rdfs:subClassOf#DessertWine>;
<vg:ClassDataPropertyLink#ClassDataPropertyLink1>
    p:class                      <rdfs:class#StrawWine>;
    p:dataProperty               <owl:DataProperty#hasColor>;
    p:value                      <rdf:resource#Golden>;

<vg:ClassDataPropertyCardinality#ClassDataPropertyCardinality1>
    p:class                      <rdfs:class#StrawWine>
    p:dataProperty               <owl:DataProperty#hasBody>
    p:value                      <rdf:resource#Full> and <rdf:resource#Moderate>

<vg:ClassDataPropertyCardinality#ClassDataPropertyCardinality2>
    p:class                      <rdfs:class#StrawWine>
    p:dataProperty               <owl:DataProperty#madeFromGrape>
    p:value  ((<rdf:resource#CabernetSauvignon> and <rdf:resource#Carbernetfranc>)
         or (<rdf:resource#Chardonnay> and <rdf:resource#SauvignonBlanc>))
```

Script 3. Version graph extended to include description og new object properties

```
# VersionGraph2 description
<vg:VersionGraph#VersionGraph2>
    p:hasPreviousVersionGraph    <vg:VersionGraph#VersionGraph1>;
    p:hasDate                    "12/05/2010";
    p:hasAuthor                  "Perrine PITTET";
    p:hasInstanceVersionGraph    <vg:InstanceVersionGraph#InstanceVersionGraph1>;

# AssociatedInstanceVersionGraph1 description
<vg:InstanceVersionGraph#InstanceVersionGraph1>
    p:hasAddIndividual           <vg:AddIndividual#AddIndividual1>
    p:hasAddMemberClass          <vg:AddMemberClass#AddMemberClass1>
    p:hasAddObjectPropertyAssertion
            <vg:AddObjectPropertyAssertion#AddObjectPropertyAssertion1>

# InstanceOperationdescription
```

```
<vg:AddIndividual#AddIndividual1>
    p:individual                <rdf:resource#VinPaillé>

<vg:AddMemberClass#AddMemberClass1>
    p:individual                <rdf:resource#VinPaillé>
    p:class                     <rdfs:class#StrawWine>

<vg:AddObjectPropertyAssertion#AddObjectPropertyAssertion1>
    p:individual                <rdf:resource#VinPaillé>
    p:objectProperty            <owl:ObjectProperty#locatedIn>
    p:value                     <rdf:resource#FrenchRegion>
```

Then, we want to add an individual of Straw Wine type: "Vin Paillé de Corrèze". First, we need to validate the previous changes by a "Commit". Then changes in the schema are recorded and the new schema version is propagated to the ontology. A third `VersionGraph`is generated for the addition of the individual. This time it contains an `InstanceVersionGraph` (Script 3).

## 6 CONCLUSION

Ontology evolution and versioning are recent domains of search. Most of the current ontology versioning approaches are not based on the evolution process. Rare are the solutions which integrate these mechanisms since the creation of the ontology. Our proposed architecture Versiongraph is a semantic solution towards the characterization of a dynamic ontology which reaches these objectives. Our ongoing research shows preliminary results on evolution of several ontologies like Wine. The architecture is employed to guide the ontology change validation in a systematic and optimized way, reducing user dependency and justifying change costs. Our short coming plan is to enhance our evolution and versioning process on several projects applied to online press comments, tourism and town heritage ontologies. Currently, we work on enlarging the set of considered OWL ontology changes and analyzing the semantic of consistency resolution of those changes to define more resolution patterns.